\documentclass[letter]{aa} 
\usepackage{graphicx}
\usepackage{txfonts}
\usepackage[authoryear]{natbib}
\bibpunct{(}{)}{;}{a}{}{,}
\begin{document}

\title{Revealing the extended radio emission from \\the gamma-ray binary HESS~J0632+057}


\author{J. Mold\'on, M. Rib\'o, J.M. Paredes}
\institute{Departament d'Astronomia i Meteorologia, Institut de Ci\`encies del
Cosmos (ICC), Universitat de Barcelona (IEEC-UB), Mart\'{\i} i Franqu\`es 1,
08028 Barcelona, Spain\\
\email{jmoldon@am.ub.es; mribo@am.ub.es; jmparedes@ub.edu}}

\authorrunning{Mold\'on et~al.}
\titlerunning{Extended radio emission of HESS~J0632+057}

\date{Received / Accepted}

\abstract
{After the detection of a 321-day periodicity in X-rays, \object{HESS~J0632+057} can be robustly considered a new member of the selected group of gamma-ray binaries. These sources are known to show extended radio structure on scales of milliarcseconds (mas).}
{We present the expected extended radio emission on mas scales from \object{HESS~J0632+057}.}
{We observed \object{HESS~J0632+057} with the European VLBI Network (EVN) at 1.6~GHz in two epochs: during the January/February 2011 X-ray outburst and 30 days later.}
{The VLBI image obtained during the outburst shows a compact $\sim$0.4~mJy radio source, whereas 30~days later the source has faded and appears extended, with a projected size of $\sim$75~AU. The peak of the emission is displaced between runs 21$\pm$5~AU, which is bigger than the orbit size. The position of the radio source is compatible with the Be star \object{MWC~148}, which sets the proper motion of the binary system below 3~mas~yr$^{-1}$ in each coordinate. The brightness temperature of the source is above $2\times10^{6}$~K. We compare the multiwavelength properties of \object{HESS~J0632+057} with those of the previously known gamma-ray binaries.}
{\object{HESS~J0632+057} displays extended and variable non-thermal radio emission. Its morphology, size, and displacement on AU scales are similar to those found in the other gamma-ray binaries, \object{PSR~B1259$-$63}, \object{LS~5039}, and \object{LS~I~+61~303}, supporting a similar nature for \object{HESS~J0632+057}.}


\maketitle

\section{Introduction} \label{introduction}

Gamma-ray binaries are binary systems containing a compact object orbiting a companion star, which are able to produce GeV and/or TeV emission and show the peak of the spectral energy distribution (SED) at MeV-GeV energies. The three ``classical'' gamma-ray binaries are \object{PSR~B1259$-$63}, \object{LS~5039}, and \object{LS~I~+61~303}, which have been unambiguously detected as point-like sources at TeV energies \citep[see the reviews in][]{paredes08,bosch11}. The broadband emission from radio to very high energy (VHE) gamma rays of these sources is variable and periodic, following the orbital period of the binary system. These three sources present variable milliarcsecond (mas) scale radio structures that have been detected with Very Long Baseline Interferometry (VLBI). We note that other binary systems have been detected up to GeV energies, but not confirmed at TeV energies \citep[e.g., ][for Cygnus~X-3]{tavani09, abdo09}, including the gamma-ray binary candidate \object{1FGL~J1018.6$-$5856} \citep{corbet11,pavlov11}. Two main scenarios have been proposed to understand the multiwavelength behaviour of gamma-ray binaries, the basic difference being the nature of the particle accelerator. In one of them, particles are accelerated in the jets of a microquasar \citep[see, e.g.,][]{bosch09}. In the other one, particles are accelerated in the shock between the relativistic wind of a young non-accreting pulsar and the stellar wind of the massive companion star \citep[see][]{tavani97,dubus06}.

\object{HESS~J0632+057} is the most recent addition to the selected group of gamma-ray binaries. The HESS Collaboration found the point-like TeV source \object{HESS~J0632+057} \citep{aharonian07}, which was variable at TeV energies \citep{acciari09}. The source has a variable X-ray counterpart \citep{hinton09,acciari09}, as well as a variable radio counterpart \citep{skilton09}. The massive B0pe star \object{MWC~148} (\object{HD~259440}), located at $\sim$1.5~kpc, was proposed as the optical counterpart \citep{hinton09}. The SED from radio to TeV energies is very similar to the one of \object{LS~I~+61~303}, which also contains a Be star \citep{hinton09}. All these results suggest that \object{HESS~J0632+057} is a new gamma-ray binary, displaying an SED one order of magnitude fainter than \object{LS~I~+61~303}. This is an important fact that could shed light on the luminosity distribution of this new population of binary systems in the Galaxy.

Recently, a periodicity of $321\pm5$~d has been revealed thanks to long-term X-ray observations conducted with {\it Swift}/XRT, strongly supporting the binary nature of the source \citep{bongiorno11}. This orbital period is between the period of \object{PSR~B1259$-$63} (1237~days) and the one of \object{LS~I~+61~303} (26.5~days). The X-ray light curve, which covers three orbital cycles, shows a bright peak and a fainter secondary peak separated $\sim$0.5 in phase. On the other hand, optical spectroscopic observations have not allowed the binary nature of the source to be unveiled up to now \citep{aragona10,casares11}.

The last bright X-ray outburst of \object{HESS~J0632+057} occurred in February 2011. \cite{falcone11} report increased X-ray activity detected by {\it Swift}/XRT between January 23 and February 6, 2011 (MJD~55584--55598). The VERITAS Collaboration reports increased activity at energies above 300~GeV between February 7 and 8, 2011 (MJD~55599--55600) \citep{ong11}. The MAGIC Collaboration reported increased gamma-ray flux above 200~GeV during February 7--9, 2011 (MJD~55599--55601) \citep{mariotti11}, confirming the VHE active state and lowering the measured energy threshold. During the outburst, we observed the source with the European VLBI Network (EVN) to explore its radio emission on mas scales.

The non-thermal radio emission produced by gamma-ray binaries has been observed on scales of 1--100~mas. \cite{massi04} report a jet-like and precessing structure with an extension of 10--80~mas (25--200~AU) from \object{LS~I~+61~303}. VLBA observations at 8~GHz along the orbital cycle \citep{dhawan06} have shown that the $\sim$7~mas (14~AU) structure follows a cometary behaviour, while the peak of the emission traces approximately an ellipse. For \object{LS~5039}, extended radio emission has been found at 5~GHz, with projected sizes of 6~mas (15~AU) with the VLBA \citep{paredes00, ribo08}, and between 60 and 300~mas (150--750~AU) with the EVN and MERLIN \citep{paredes02}. Observations with the VLBA covering an orbital cycle show periodic morphological variability and hints of peak displacement \citep{moldon11c}. Recently, \cite{moldon11a} has shown that \object{PSR~B1259$-$63}, the only gamma-ray binary that is known to contain a pulsar, displays emission with a projected extension of 50~mas (120~AU) at 2.3~GHz during periastron. The peak of the radio nebula is detected at distances between 10 and 50~AU from the system.

Here we present the first VLBI radio images of \object{HESS~J0632+057}, obtained with the EVN during and after the high-energy activity period in early 2011 \citep[preliminary results were presented in][]{moldon11b}. The observations and data reduction are described in Sect.~\ref{observations}. The results can be found in Sect.~\ref{results}. A discussion of these results in the context of gamma-ray binaries can be found in Sect.~\ref{discussion}.

\section{Observations and data reduction} \label{observations}

\begin{figure*}[] 
\begin{center}
\resizebox{0.73\hsize}{!}{\includegraphics[angle=0]{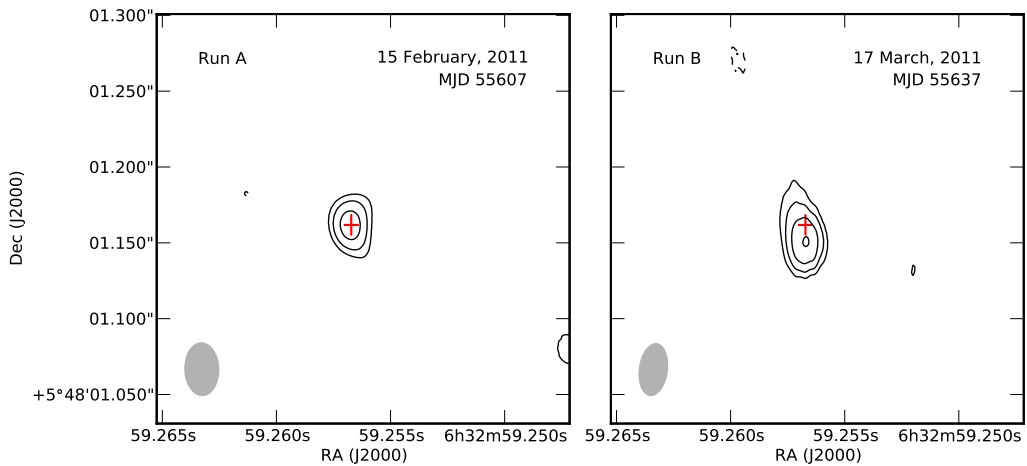}}
\caption{EVN radio images of \object{HESS~J0632+057} at 1.6~GHz during the X-ray outburst and 30~days later. North is up and east to the left. The observation dates are quoted at the top of each panel. The synthesized beam (see Table~\ref{table:parameters}) is displayed in the bottom-left corner of each image. The red crosses mark the position and 3$\sigma$ uncertainty of the fitted component of run~A. For each image, the displayed contours start at 3$\sigma$ and increase by factors of $2^{1/2}$; the 1$\sigma$ rms close to the source
is 50~$\mu$Jy~beam$^{-1}$ in run~A and 13~$\mu$Jy~beam$^{-1}$ in run~B.
\label{fig:fig1}}
\end{center}
\end{figure*}

Following the report of X-ray activity from \object{HESS~J0632+057} between January 23 and February 6, 2011 \citep{falcone11}, we observed the source with the EVN in target of opportunity (ToO) mode in two epochs separated by 30~days. The first radio continuum observation (run~A) was conducted at 1.6~GHz (18~cm) during eight hours on February 15, 2011 (MJD~55607), UTC~15:50 to 00:00. The antenna network consisted in seven antennas: Effelsberg, Jodrell Bank Lovell, Medicina, Onsala, Torun, Westerbork, and Hartebeesthoek, providing baselines ranging from 200 to 8\,000~km (200 to 1\,000~km without Hartebeesthoek). A data rate of 1024~Mbps per station was directly streamed to the central processor at JIVE and correlated in real time (e-VLBI). The rapid response of the e-EVN observations and correlation allowed us to perform a second ToO (run~B) in a disk-recorded session with 12 EVN antennas: Effelsberg, Jodrell Bank Lovell, Medicina, Onsala, Torun, Westerbork, Hartebeesthoek, Svetloe, Badary, Zelenchukskaya, Nanshan (Urumqi), and Sheshan (Shanghai), providing baselines ranging from 200 to 10\,000~km. The observation lasted ten hours and was conducted on March 17, 2011 (MJD~55637), UTC~13:00 to 23:00, with a data rate of 1024~Mbps. This second epoch provides considerably better sensitivity. The EVN project codes for the observations are RR005 and RM006, respectively.

Both observations have a similar structure, switching between HESS~J0632+057 and the phase reference calibrator \object{J0619+0736}, separated 3\fdg9, with cycling times of five minutes to avoid losing phase coherence. The phase reference calibrator was correlated at the position $\alpha_{\rm J2000.0}=06^{\rm h} 19^{\rm m} 09\fs9710$ and $\delta_{\rm J2000.0}=07\degr 36\arcmin 41\farcs220$ in the frame of ICRF (Goddard VLBI global solution 2010a\footnote{http://gemini.gsfc.nasa.gov/solutions/2010a/2010a.html}). This position has absolute uncertainties of 2.0 and 1.1~mas in right ascension and declination, respectively. The fringe finders DA193, 3C147, and OQ208 were observed during run~A and 0528+134, and DA193 during run~B.

The data reduction was performed in AIPS\footnote{The NRAO Astronomical Image Processing System. http://www.aips.nrao.edu/}. Flagging based on predicted off-source times, owing to slewing or failures, was applied using UVFLG. A priori visibility amplitude calibration used the antenna gains and the system temperatures measured at each station. The amplitude calibration was improved by scaling the individual antenna gains by a factor obtained from the phase calibrator and fringe-finder models. We used ionospheric total electron content (TEC) models based on GPS data obtained from the CDDIS data archive\footnote{The Crustal Dynamics Data Information System http://cddis.nasa.gov/} to correct the global phase variations due to the ionosphere. The parallactic angle correction was applied with VLBAPANG. Fringe fitting on the phase reference calibrator was performed with the AIPS task FRING, and the solutions were applied to the target source. The instrumental bandpasses were corrected using BPASS. The data were averaged in frequency and time, and clean images were produced with IMAGR. A cell size of 1~mas was used for cleaning both epochs. The images were produced using a weighting scheme with robust parameter 0 for run~A, and 2 (which is slightly more sensitive to extended emission) for run~B. A tapering of 30~M$\lambda$ was applied to avoid the presence of possible unreliable high-resolution features due to sidelobes of the synthesized beam. No self-calibration of the data was possible because of the low flux density of the target source.

\begin{table*} 
\begin{center}
\caption{Observation parameters and Gaussian components fitted to the images.}
\label{table:parameters} 
\begin{tabular}{llccrlr@{ $\pm$ }l@{}r@{ $\pm$ }lr@{ $\pm$ }lr@{ $\pm$ }lr@{ $\pm$ }lr@{ $\pm$ }l c}
\hline
\hline
Run & MJD & Num. & $\theta_{\rm HPBW}$ & $\theta_{\rm PA}$ & Comp. & \multicolumn{2}{c}{$S_{\rm peak}$} & \multicolumn{2}{c}{$S_{\rm 1.6~GHz}$} & \multicolumn{2}{c}{$\Delta\alpha$\footnotemark[1]} & \multicolumn{2}{c}{$\Delta\delta$\footnotemark[1]} & \multicolumn{2}{c}{Maj. Axis} &\multicolumn{2}{c}{Min. Axis} &  Type \\

    &     &  anten.         & [${\rm mas^{2}}$]   & [$\degr$]    &       & \multicolumn{2}{c}{[${\rm \mu Jy~b^{-1}}$]}      & \multicolumn{2}{c}{[${\rm \mu Jy}$]}           & \multicolumn{2}{c}{[mas]} &          \multicolumn{2}{c}{[mas]}          & \multicolumn{2}{c}{[mas]}     &\multicolumn{2}{c}{[mas]}    &  \\
\hline
A  & 55607.83 &  7 & 36$\times$23 & 2    & Core1 & 340 & 50 & 410 & 90 & \multicolumn{2}{c}{---} & \multicolumn{2}{c}{---} & 12 & 13 & 12 & 10 &  Point-like\\
\hline
B  & 55637.75 & 12 & 35$\times$19 & $-$7 & Core2    & 81 & 14 &  90 & 25 & $-$3 & 2 & $-$14 & 3 & 9  & 11 & 12 & 12 &  Point-like\\
   &          &    &              &      & Extended & 56 & 13 & 110 & 40 &    6 & 3 &     5 & 6 & 48 & 18 & 11 & 10 &  Extended  \\
\hline
\end{tabular}
\end{center}
$^{a}$Displacements with respect to the red crosses in Fig.~\ref{fig:fig1}.
\end{table*}

\section{Results} \label{results}

The resulting VLBI images at 1.6~GHz are shown in Fig.~\ref{fig:fig1}. Run~A shows a $410\pm90~\mu$Jy point-like source. Run~B, observed 30~days later, shows a $180\pm30~\mu$Jy (measured with TVSTAT within AIPS) faint source with extended emission towards north. The 1$\sigma$ rms noise close to the source is 50~$\mu$Jy~beam$^{-1}$ and 13~$\mu$Jy~beam$^{-1}$, respectively. We note that the flux density of the extended emission detected in run~B is below the rms noise limit of the image of run~A. In Table~\ref{table:parameters} we show some parameters of the observations, the obtained synthesized beam, and the fitted components to the source (Gaussian components obtained with JMFIT within AIPS).

The emission in run~A is described by a component located at $\alpha_{\rm J2000.0}=06^{\rm h} 32^{\rm m} 59\fs2567(1)$ and $\delta_{\rm J2000.0}=05\degr 48\arcmin 01\farcs162(2)$. The values in parenthesis refer to the uncertainty in the last digit (1.3 and 2~mas, respectively), and correspond to the formal errors of the fit. Run~B is described well by a point-like core (see Table~\ref{table:parameters}) and an extended component (which has an intrinsic PA of $4\pm20\degr$) located at 21$\pm$6~mas in PA $30\pm10\degr$ with respect to Core2. The total extension of the source is $\sim$50~mas.

The relative astrometry shows that the peak of the emission suffers a total displacement of 14$\pm3$~mas in PA $190\pm9\degr$ between runs~A and B. This displacement is aligned with the direction of the extended emission. The separation between the peak positions corresponds to a projected linear distance of 21$\pm$5~AU. Covering this distance in the 30~days between runs requires a constant velocity of $1200\pm300$~km~s$^{-1}$.

\section{Discussion} \label{discussion}

The first VLBI image of \object{HESS~J0632+057} at 1.6~GHz, taken a few days after the peak of the January 2011 X-ray outburst, shows a compact $410~\mu$Jy point-like source (see Fig.~\ref{fig:fig1}). After 30 days, the source displays one-sided extended emission with a total flux density of $180~\mu$Jy and with an extension of 50~mas (75~AU assuming a distance to the system of 1.5~kpc) in PA $30\degr$. The peak of the emission is displaced 14~mas (21~AU) in PA $190\degr$. We note that the rms noise of the first image is above the flux density of the extended emission in the second image. 

Even though the orbital parameters of \object{HESS~J0632+057} are unknown, the size of the orbit can be derived from the orbital period (321~days), assuming a mass of the compact object of 1.4~$M_\odot$, and a Be star of 16$\pm3~M_\odot$ \citep{aragona10}. The semimajor axis of the orbit is 2.4$\pm$0.1~AU, which corresponds to a projected distance of 1.6$\pm$0.1~mas (or less considering the inclination and the argument of the periastron of the orbit). The peak displacement reported here, 14~mas, cannot be explained by any motion within the orbit.

The measured position in run~A is clearly compatible with the UCAC3 catalogue position of the Be star \object{MWC~148}, which has an uncertainty of 14~mas in each coordinate \citep{zacharias10}, therefore the detected radio source is unambiguously related to the Be star. We computed the proper motion of the source using the position from run~A with an uncertainty of 25~mas to include possible peak displacements, the position of \object{MWC~148} from UCAC3, obtained at mean epoch 1984.2 \citep{zacharias10}, and the VLA-D and GMRT positions from \citet{skilton09} (with uncertainties of 300 and 500~mas, respectively). Despite the time span of 27 years of observations, the proper motion is compatible with no motion. The result is $\mu_{\alpha} \cos(\delta) = 0.2\pm1.0$~mas~yr$^{-1}$ and $\mu_{\delta} = -0.2\pm0.9$~mas~yr$^{-1}$.

The brightness temperature corresponding to the radio source in run~A (see Table~\ref{table:parameters}) at 1.6~GHz is $2\times10^{6}$~K. Such a high value, together with non-simultaneous radio observations suggesting a negative spectral index around $\sim-0.6$ \citep{skilton09}, rules out a thermal emission mechanism. Non-thermal synchrotron radiation remains, therefore, as the most plausible interpretation for the \object{HESS~J0632+057} radio emission. In this case, and assuming  equipartition between the relativistic electrons and the magnetic field, the observed properties correspond to a total energy between 0.1 and 100~GHz of 4$\times10^{38}$~erg, with an equipartition magnetic field of $\sim$0.02~G. These values are approximately ten times lower than those obtained for \object{LS~5039} from VLBI observations \citep{paredes00}.

The gamma-ray binaries \object{PSR~B1259$-$63}, \object{LS~5039}, and \object{LS~I~+61~303} show a radio morphology with a central core and one-sided extended radio emission on scales of a few AU (although bipolar extended emission has also been detected at some orbital phases: \citealt{dhawan06, ribo08, moldon11a, moldon11c}). In all these cases, the authors find morphological changes on timescales close to the orbital period, and displacements of the peak of the emission bigger than the orbit size. The morphology and peak displacement reported here show that the radio emission of \object{HESS~J0632+057} is similar to the ones observed in the other known gamma-ray binaries, suggesting a similar nature for \object{HESS~J0632+057}.

\begin{figure}[t!!] 
\resizebox{1.0\hsize}{!}{\includegraphics[angle=0]{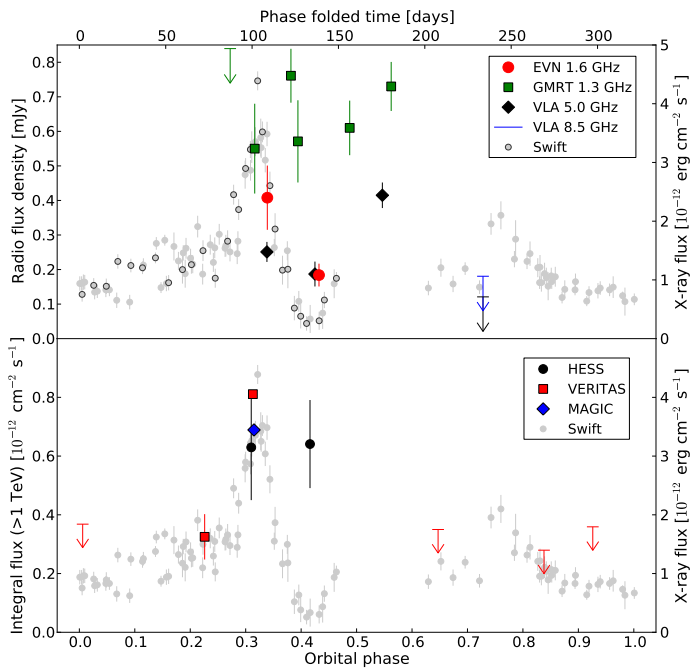}}
\caption{Multiwavelength light curve of \object{HESS~J0632+057}. The data were folded with a period of 321~days. Zero phase has been arbitrarily defined at MJD~54857. The grey circles are X-ray observations by {\it Swift}/XRT (0.3--10~keV) from \cite{bongiorno11}. {\it Top}: the red circles are the EVN observations from this paper. The other radio data are from \citet{skilton09}. The black circles (filled in grey) are {\it Swift}/XRT data obtained during the same cycle as the EVN data. {\it Bottom}: TeV emission from \citet{acciari09,ong11,mariotti11,mukherjee11}.
\label{fig:fig2}}
\end{figure}

In Fig.~\ref{fig:fig2} we show the phase-folded light curve of the X-ray data from \cite{bongiorno11}, together with radio (top panel) and TeV (bottom panel) data. Our two EVN measurements suggest a radio flux decrease of the most compact region of the source similar to the X-ray one after the outburst, although at a slower rate. However, a VLBI monitoring of the outburst is required to understand the flux changes better. On the other hand, a persistent and higher flux was measured in previous lower resolution GMRT observations at a similar frequency up to $\sim$100~days after the outburst. This can be explained by the presence of diffuse (resolved) emission, which is lost on VLBI scales (the shortest significant baseline in our data samples spacial frequencies around 1~M$\lambda$, which corresponds to angular scales of 200 mas).

\object{LS~I~+61~303} shows an X-ray outburst before apastron that is correlated with the TeV outburst \citep{anderhub09, zabalza11}. The duration of the X-ray outburst of \object{HESS~J0632+057} is $\sim$5 times longer (although $\sim$0.1 in phase in both cases). We computed non-simultaneous TeV/X-ray flux ratios during the 2011 outburst (around phase 0.31) and compared them with the ones of \object{LS~I~+61~303}. The TeV/X-ray flux ratio of \object{HESS~J0632+057} is $\sim$5 higher (the X-ray flux is lower, or the TeV emission is higher) than the one of \object{LS~I~+61~303}. This is consistent with a lower magnetic field in \object{HESS~J0632+057}. On the other hand, TeV emission has been detected at similar levels during the X-ray dip, and thus does not follow the X-ray light curve (Fig.~\ref{fig:fig2}). When comparing the average radio (0.7~mJy) to X-ray flux ratios, we see that the X-ray flux is higher, or the radio emission is lower, in \object{HESS~J0632+057} by a factor $\sim$30. The outbursts from these two sources are observationally very different, although the lack of information regarding orbital parameters prevents any further discussion.

As a final comment, we could interpret the displacement between the peaks of runs~A and B (14~mas in 30~days) as produced by the proper motion of a blob ejected within a microquasar (see \citealt{mirabel94} for \object{GRS~1915+105}). A blob velocity above 0.1c requires that the jet is pointing towards the observer with an angle below 2$^{\circ}$, which would be either very restrictive or fine tuned.

In conclusion, \object{HESS~J0632+057} displays extended and variable radio emission at 50--100~AU scales, with a projected displacement of the peak of the emission of 21~AU in 30~days. Similar morphologies and behaviours have been found in the other gamma-ray binaries. However, a more detailed monitoring of the variability of the source with high-sensitivity VLBI observations along the orbital cycle is required to measure the total extension and morphology of the extended emission, to measure the morphologic and astrometric changes at different orbital phases, and to search for periodicity in these changes.

\begin{acknowledgements}
We thank the EVN PC Chair, Tiziana Venturi, for supporting our ToO observations and to the EVN stations who made this possible. e-VLBI developments in Europe are supported by NEXPReS, an Integrated Infrastructure Initiative (I3), funded by the European Union Seventh Framework Programme (FP7/2007-2013) under grant agreement RI-261525. The European VLBI Network (http://www.evlbi.org/) is a joint facility of European, Chinese, South African, and other radio astronomy institutes funded by their national research councils. We acknowledge support by the Spanish Ministerio de Ciencia e Innovaci\'on (MICINN) under grants AYA2010-21782-C03-01 and FPA2010-22056-C06-02.
J.M. acknowledges support by MICINN under grant BES-2008-004564.
M.R. acknowledges financial support from MICINN and European Social Funds through a \emph{Ram\'on y Cajal} fellowship.
J.M.P. acknowledges financial support from ICREA Academia.
\end{acknowledgements}

\bibliographystyle{aa}
\bibliography{art}

\end{document}